\documentclass[%
 aip,
 amsmath,amssymb,
 reprint,%
]{revtex4-1}

\usepackage{graphicx}% Include figure files
\usepackage{dcolumn}% Align table columns on decimal point
\usepackage{bm}% bold math
\usepackage[utf8]{inputenc}
\usepackage[T1]{fontenc}
\usepackage{mathptmx}
\usepackage{etoolbox}

\usepackage{float}
\usepackage{placeins}

\usepackage{xcolor,soul}
\sethlcolor{yellow!30} %for color  \hl{text}

\makeatletter
\def\@email#1#2{%
 \endgroup
 \patchcmd{\titleblock@produce}
  {\frontmatter@RRAPformat}
  {\frontmatter@RRAPformat{\produce@RRAP{*#1\href{mailto:#2}{#2}}}\frontmatter@RRAPformat}
  {}{}
}%
\makeatother
\begin{document}

\preprint{AIP/123-QED}

\title{Development of a compact cryogenic Penning trap with permanent magnets: An intermediate step toward the Shanghai Penning Trap}
% Force line breaks with \\
\author{Tianhang Zhang}
%\thanks{}
\affiliation{The Key Laboratory of Nuclear Physics and Ion Beam Applications of the Ministry of Education, Department of Nuclear Science and Technology, the Institute of Modern Physics, Fudan University, Shanghai, 200433, China}

\author{Jiawei Wang}
\affiliation{The Key Laboratory of Nuclear Physics and Ion Beam Applications of the Ministry of Education, Department of Nuclear Science and Technology, the Institute of Modern Physics, Fudan University, Shanghai, 200433, China}

\author{Jialin Liu}
\affiliation{The Key Laboratory of Nuclear Physics and Ion Beam Applications of the Ministry of Education, Department of Nuclear Science and Technology, the Institute of Modern Physics, Fudan University, Shanghai, 200433, China}

\author{Jingtian Wei}
\affiliation{Department of Physics, University of Toronto, 60 St George St., Toronto, M5S 1A7 Ontario, Canada}

\author{Jiaxuan Ji}
\affiliation{The Key Laboratory of Nuclear Physics and Ion Beam Applications of the Ministry of Education, Department of Nuclear Science and Technology, the Institute of Modern Physics, Fudan University, Shanghai, 200433, China}

\author{Jifei Wu}
\affiliation{The Key Laboratory of Nuclear Physics and Ion Beam Applications of the Ministry of Education, Department of Nuclear Science and Technology, the Institute of Modern Physics, Fudan University, Shanghai, 200433, China}

\author{Zichen Su}
\affiliation{The Key Laboratory of Nuclear Physics and Ion Beam Applications of the Ministry of Education, Department of Nuclear Science and Technology, the Institute of Modern Physics, Fudan University, Shanghai, 200433, China}

\author{Yiming Xie}
\affiliation{The Key Laboratory of Nuclear Physics and Ion Beam Applications of the Ministry of Education, Department of Nuclear Science and Technology, the Institute of Modern Physics, Fudan University, Shanghai, 200433, China}

\author{Liangyu Huang}
\affiliation{The Key Laboratory of Nuclear Physics and Ion Beam Applications of the Ministry of Education, Department of Nuclear Science and Technology, the Institute of Modern Physics, Fudan University, Shanghai, 200433, China}

\author{Ke Yao}
\affiliation{The Key Laboratory of Nuclear Physics and Ion Beam Applications of the Ministry of Education, Department of Nuclear Science and Technology, the Institute of Modern Physics, Fudan University, Shanghai, 200433, China}

\author{Yang Shen}
\affiliation{The Key Laboratory of Nuclear Physics and Ion Beam Applications of the Ministry of Education, Department of Nuclear Science and Technology, the Institute of Modern Physics, Fudan University, Shanghai, 200433, China}

\author{Yaming Zou}
\affiliation{The Key Laboratory of Nuclear Physics and Ion Beam Applications of the Ministry of Education, Department of Nuclear Science and Technology, the Institute of Modern Physics, Fudan University, Shanghai, 200433, China}

\author{Baoren Wei}
\affiliation{The Key Laboratory of Nuclear Physics and Ion Beam Applications of the Ministry of Education, Department of Nuclear Science and Technology, the Institute of Modern Physics, Fudan University, Shanghai, 200433, China}

\author{Bingsheng Tu$^{\ast}$}
\affiliation{The Key Laboratory of Nuclear Physics and Ion Beam Applications of the Ministry of Education, Department of Nuclear Science and Technology, the Institute of Modern Physics, Fudan University, Shanghai, 200433, China}
\email{Corresponding author: bingshengtu@fudan.edu.cn}

\date{\today}% It is always \today, today,

\begin{abstract}
Penning traps, renowned for their unparalleled precision in determining fundamental properties such as mass and magnetic moments, are cornerstone instruments in modern physics. Their applications span from nuclear structure studies to stringent tests of quantum electrodynamics and CPT invariance. Although Penning traps have been demonstrated for fundamental studies, often employing superconducting magnets, their high cost and operational complexity remain challenges. In this work, we report the development of a compact cryogenic Penning trap that utilizes a permanent magnet to provide a confining magnetic field, offering a more economical and flexible alternative. We have successfully demonstrated all core functionalities of this system, including ion generation, transport, confinement, manipulation, and signal detection. This compact trap not only serves as a vital technical testbed for the development of the Shanghai Penning Trap, but also establishes a cryogenic Penning-trap experiment platform for ion trapping and cooling applications as well as envisaged spectroscopic studies applications.
\end{abstract}

\maketitle

\section{\label{sec:introduction} INTRODUCTION}

The Penning trap's unparalleled precision in determining properties like mass and magnetic moments have driven advances across modern physics\cite{Dilling2018, Blaum2010, Blaum2021}. In nuclear structure physics, Penning trap mass spectrometry provides definitive mass values far from stability, offering crucial insights into shell evolution, the emergence of new magic numbers, and the properties of exotic nuclear systems\cite{Block2010, Orford2018, Ramirez2012} . Beyond nuclear science, Penning traps serve as powerful laboratories for exploring fundamental physics. They enable stringent tests of quantum electrodynamics (QED) in strong fields through precision measurements of the electron g factor in highly charged ions \cite{Morgner2023, Morgner20250328, Morgner20250529}, and provide the most precise comparisons of particles and antiparticles to test CPT invariance \cite{Borchert2022, Smorra2017, Anderson2023}.

In this work, we report the development of a compact cryogenic Penning trap, which employs a permanent magnet to provide the confining magnetic field. Within this system, we have successfully demonstrated ion generation, transport, confinement, manipulation and signal detection. This compact trap not only serves as a technical testbed for the development of the Shanghai Penning Trap for metastable states studies \cite{Tu2023,Wang2025}, but also provides essential groundwork for advanced ion trapping applications. Transport of trapped ions over long distances represents a promising application with such a portable design. Previous studies have demonstrated the feasibility of relocating charged particles using Penning traps. For example, Tseng and Gabrielse transported electrons over 5000 km using a Penning trap in a superconducting magnet, maintaining confinement under cryogenic conditions without external power \cite{Tseng1993}. More recently, Leonhardt et al. successfully relocated trapped protons across the CERN site using the transportable BASE-STEP apparatus, designed for precision antimatter experiments \cite{Leonhardt2025}. Among these instruments, the use of superconducting magnets remains costly and carries the risk of quenching, whereas permanent magnets are entirely free from this issue. A more economical approach is to employ Penning trap technology based on cryogenic permanent magnets. Although the compact cryogenic permanent magnet Penning trap is constrained by magnetic field homogeneity and stability for high-precision mass measurements, it provides a solid electro-magnetic field environment to store different types of ions. In addition, it offers a promising platform for laser cooling and spectroscopic studies based on trapped ions in the cryogenic environment. For instance, J. M. Cornejo et al. demonstrated laser cooling of Be ions \cite{Cornejo2023,Cornejo2024} and motivated towards the sympathetic cooling\cite{Niemann2019}. S. H. Hoogerheide et al. employed a room-temperature permanent magnet Penning trap for laser spectroscopy of highly charged ions \cite{Hoogerheide2015} and successfully determined the life time of Ar$^{9+}$\cite{PhysRevA.98.032501}. With the ion trapping and manipulation techniques demonstrated in the compact cryogenic Penning trap, we can envisage future applications such as laser cooling and spectroscopy with this system.

\section{\label{sec:experiment design} EXPERIMENT DESIGN}

\subsection{\label{sec:principle}Principle}
The operating principle of Penning traps has been comprehensively described in previous works \cite{Brown1982,Sturm2019,Heisse2019,Tu2023}. A homogeneous magnetic field coupled to an electrostatic quadrupole potential constitutes the core field configuration of a Penning trap. The ideal harmonic trapping electrostatic potential $U$ in cylindrical coordinates $(z,\rho)$ is:
\begin{equation}
U(z,\rho)=\frac{U_0C_2}{2d_{\mathrm{char}}^2}(z^2-\frac{\rho^2}{2}),
\label{eq:debye}
\end{equation}
where $U_0$ is the voltage of the ring electrode and $C_2$ represents a dimensionless coefficient. The axial electric field component confines the ions, resulting in an axial harmonic oscillation at frequency $\nu_z$. The radial electric field component, though divergent in isolation, couples with the axial magnetic field $B_z$ to sustain ion confinement. This coupling splits the radial motion into two eigenmodes: a fast modified cyclotron motion with frequency $\nu_+$ and a slow magnetron motion with frequency $\nu_-$. The free cyclotron frequency $\nu_c=\frac{Bq}{2\pi M}$, which is determined by the charge-to-mass ratio $q/M$ , is derived from the eigenfrequencies by the invariance theorem\cite{Brown1982}:
\begin{equation}
\nu_c=\sqrt{\nu_+^2+\nu_z^2+\nu_-^2}.
\label{eq:debye}
\end{equation}

Our trap utilizes a cylindrical stack of electrodes to generate the electrostatic potential, consistent with previous designs\cite{Gabrielse1984,Tan1989,Gabrielse1989,Sikdar2013,Schneider2017,Smorra2017}. The electrostatic potential generated by this cylindrical trap may differ from an ideal quadrupole, resulting in electric field imperfections. Mathematically, the general potential in the center of the measurement trap can be characterized through a series expansion,
\begin{equation}
U(\rho,\phi)=\frac{U_0}{2}\sum^\infty_{n=0}{C_n\left(\frac{r}{d_\mathrm{char}}\right)^nP_n\left(\cos{\phi}\right)},
\label{eq:debye}
\end{equation}
where $C_n$ represents a dimensionless expansion coefficient and $P_n\left(\cos{\phi}\right)$ is the Legendre polynomial. Cylindrical Penning traps employ a five-electrode structure comprising a central ring electrode, paired correction electrodes, and paired endcap electrodes. The dimensions (length, inner radius, and electrode spacing) of the ring electrode and paired correction electrodes are provided in Table~\ref{table_parameters}. This structure generates a harmonic potential, while theoretically eliminating the terms $C_4$ and $C_6$ \cite{Kohler2015}. Our trap is designed to have $C_2/2d_{\mathrm{char}}^2=\mathrm{-0.0306~mm^{-2}}$.

\begin{table}
\caption{\label{table_parameters}Trap electrode parameters, with dimensions labeled in Figure~\ref{fig_ionG}. These geometric parameters correspond to the degrees of freedom of the trap. The tuning ratio is the voltage ratio between the correction and ring electrodes, determined through a finite-element simulation of the trap's electrostatic field.}
\begin{ruledtabular}
\begin{tabular}{cc}
Trap parameters&Design values\\
\hline
Trap radius $r_{\mathrm{0}}$& 3.500 mm\\
 Distances between electrodes $dd$& 0.140 mm\\
 Length of ring $l_{\mathrm{R}}$& 0.989 mm\\
 Length of correction $l_{\mathrm{C}}$& 2.715 mm\\
 Tuning ratio $T_\mathrm{r}$& 0.8827\\
\end{tabular}
\end{ruledtabular}
\end{table}

\subsection{\label{sec:setup}Set up}
\begin{figure*}[!htbp]
\includegraphics[width=\textwidth]{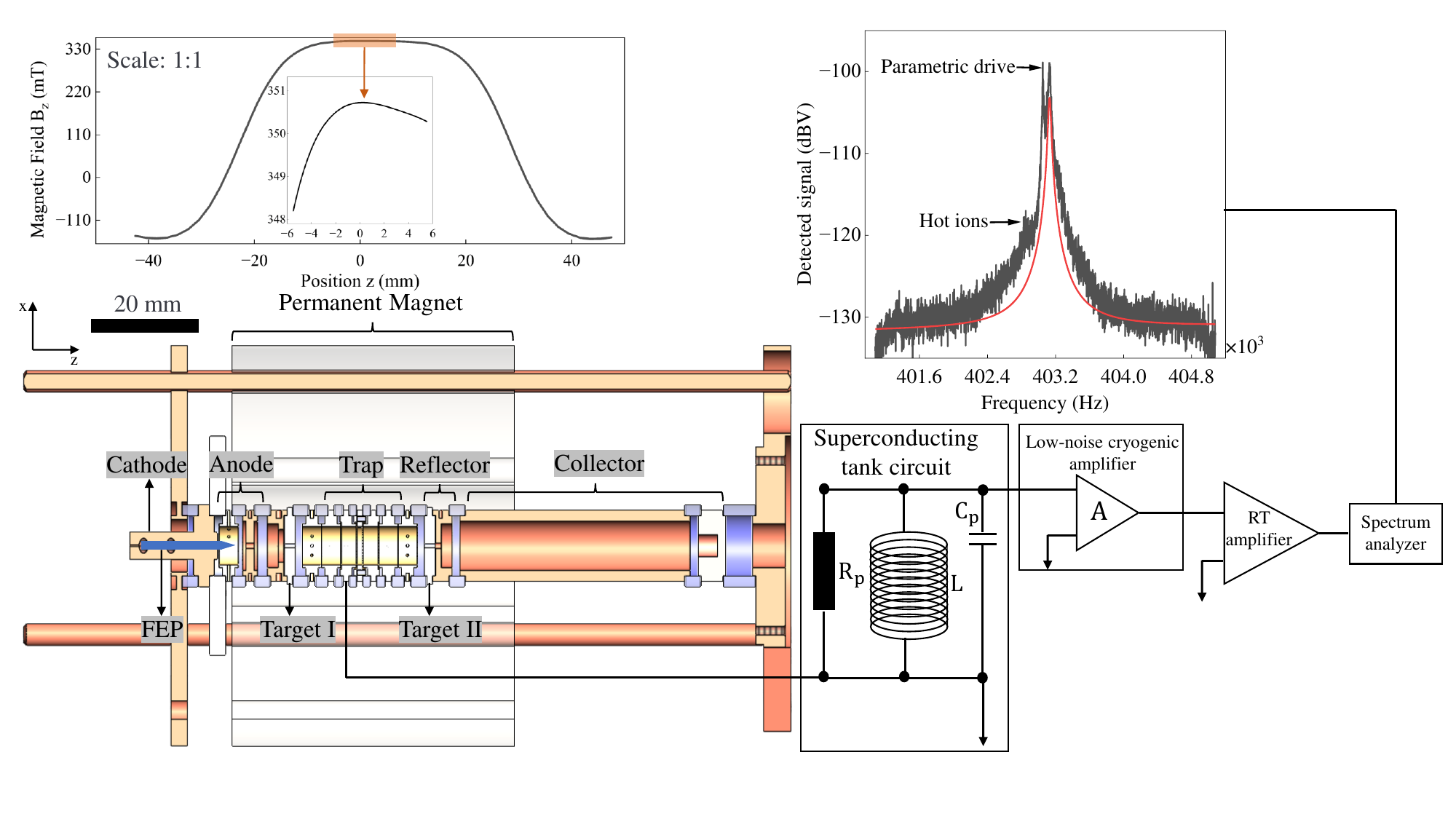}%
\caption{\label{fig_trap} Schematic of the compact cryogenic Penning trap assembly, including the detection system. The trap electrodes are segmented into the cathode, anode, trap, reflector, and collector. The gaps between electrodes are filled with sapphire insulating rings. The blue needle-like structure represents the Field Emission Point (FEP), fixed within the cathode, which is supplied with a high negative voltage. Two target electrodes (Target I and Target II), fabricated from electrically conductive carbon nanotube and featuring central apertures, are also shown. The white component surrounding the trap electrodes is a SmCo permanent magnet. The overlaid curve depicts the axial magnetic field along the trap axis, with an inset showing a magnified view of the central region. The detection system, positioned to the right of the trap, consists of a superconducting tank circuit, a low-noise cryogenic amplifier, a room-temperature (RT) amplifier, and a spectrum analyzer. The spectrum displayed adjacent to the analyzer is the ion signal excited via parametric drive. The red curve represents the baseline, obtained from a fit to the resonator's noise signal in the absence of ions. The black curve is the experimentally measured spectrum, where the peak near the resonator's resonance frequency corresponds to the parametric drive peak. The shoulder-like features on the left side of this peak are signals induced by hot ions.}
\centering
\end{figure*}

The structure of the compact cryogenic Penning trap is depicted in Figure~\ref{fig_trap}. A permanent magnet is tightly mounted around the electrode assembly, generating a uniform magnetic field of approximately 330~mT in the trap region. The magnet consists of a hollow concentric cylinder made of SmCo, with outer diameter of 73~mm and an inner diameter of 22~mm to accommodate a cylindrical electrode with outer diameters of 13~mm. It has an overall height of 49.7 mm, designed to ensure a homogeneous magnetic field at the center.Six cylindrical through-holes which are oriented parallel to the trap axis are evenly distributed at a radius of 30.3 mm in the permanent magnet, allowing it to be secured with M4 copper–beryllium rods. Compared with a superconducting magnet that can generate fields of several tesla, the permanent magnet is more compact and portable, but it produces a weaker magnetic field and exhibits less homogeneity in the central region. We measured the axial magnetic field of the permanent magnet using a Gaussmeter and observed a certain degree of asymmetry, as indicated by the magnetic field gradient in the inset of Figure~\ref{fig_trap}. To mitigate this, the center of the ion trap was positioned at the flattest region of the magnetic field variation. With the trap centered at this extreme point (z = 0 mm), the following inhomogeneity coefficients were obtained from a fit: $B_1=14.9 ~\mathrm{\mu T/mm}$, $B_2=-35.3 ~\mathrm{\mu T/mm^2}$, and $B_3=5.2 ~\mathrm{\mu T/mm^3}$. Our experiment setup features an outermost sealed chamber with a radius of 125 mm and a height of 700 mm to provide a vacuum environment. The background pressure in this outer chamber can reach down to $10^{-4}$~Pa. The two-stage cold heads (Cryomech PT415) extend vertically from top to bottom, designated as the 40~K cold head and the 4~K cold head according to their operating temperatures. The upper plate of the external chamber is equipped with four KF feedthroughs, which provide connections to external power supplies and measurement instruments. While being routed to the trap electrodes and cryogenic electronics, the DC lines are routed through three RC low-pass filter stages (filter boards) mounted in series at temperature stages of 300 K, 40 K, and 4 K. These filter boards at two cold head stages also serve as thermal anchors for the lines at their respective temperatures. The Penning trap experiment is suspended from the 4~K cold head. The entire trap assembly is enclosed in a high-purity copper chamber with a height of 109~mm and a diameter of 80~mm. Feedthroughs at the top flange of the chamber connect the cylindrical electrodes to the pins of the 4 K filter board located on the upper plate of the enclosure. The upper section of the 4~K cold head supports the cryogenic detection system including a superconducting tank circuit and a cryogenic amplifier\cite{Wang2025}.

\FloatBarrier
\FloatBarrier

The cryogenic environment is essential for this experiment, not only for operating the detection system based on the image current technique\cite{Myers2013}, but also for establishing and maintaining an ultra‑high vacuum crucial for long ion storage times. Before cooling, the outermost chamber is evacuated to a background pressure of $10^{-4}$~Pa, while the inner copper trap chamber is evacuated to below $3 \times 10^{-5}$~Pa at room temperature and then hermetically sealed (pinched off). During cooldown, cryopumping by the cold surfaces further reduces the residual gas pressure. This pinched‑off, cryogenic approach creates an extreme vacuum environment proven to enable exceptionally long confinement times. For instance, in a directly comparable setup, antiprotons were stored for 614 days within a pinched‑off chamber at ~5 K\cite{BASE2026}. In the experiment, we did not attempt to sustain ions for an extended period; instead, the ions were kicked out after being confined for a few days. Nevertheless, the achieved vacuum conditions are functionally equivalent, indicating excellent prospects for extended ion storage.

The image current, induced by ion oscillations at one trap electrode, is coupled to a superconducting tank circuit (resonator) developed in-house. For the unloaded resonator, the measured quality factor (Q value) reaches 98,004 \cite{Wang2025}. When coupled to the trap, due to the parasite capacitive loss the resonator Q value is reduced to 11,903, as determined by fitting the resonator’s noise spectrum. The resonator output signal is then sent to a low-noise cryogenic amplifier, followed by a room-temperature (RT) amplifier and a spectrum analyzer. For ion oscillation frequencies that match the resonance frequency of the superconducting tank circuit, the detection system’s impedance $Z_{\rm{LC}}$ becomes purely real, corresponding to a high parallel resistance $R_{\rm{p}}$. This impedance converts the small induced image currents into measurable voltage signals. Prior to performing a fast Fourier transform (FFT), the signal is amplified and down-converted using a local oscillator at 391 kHz.

\section{\label{sec:result} RESULT}
\begin{figure}[!htbp]
\includegraphics[width=0.5\textwidth]{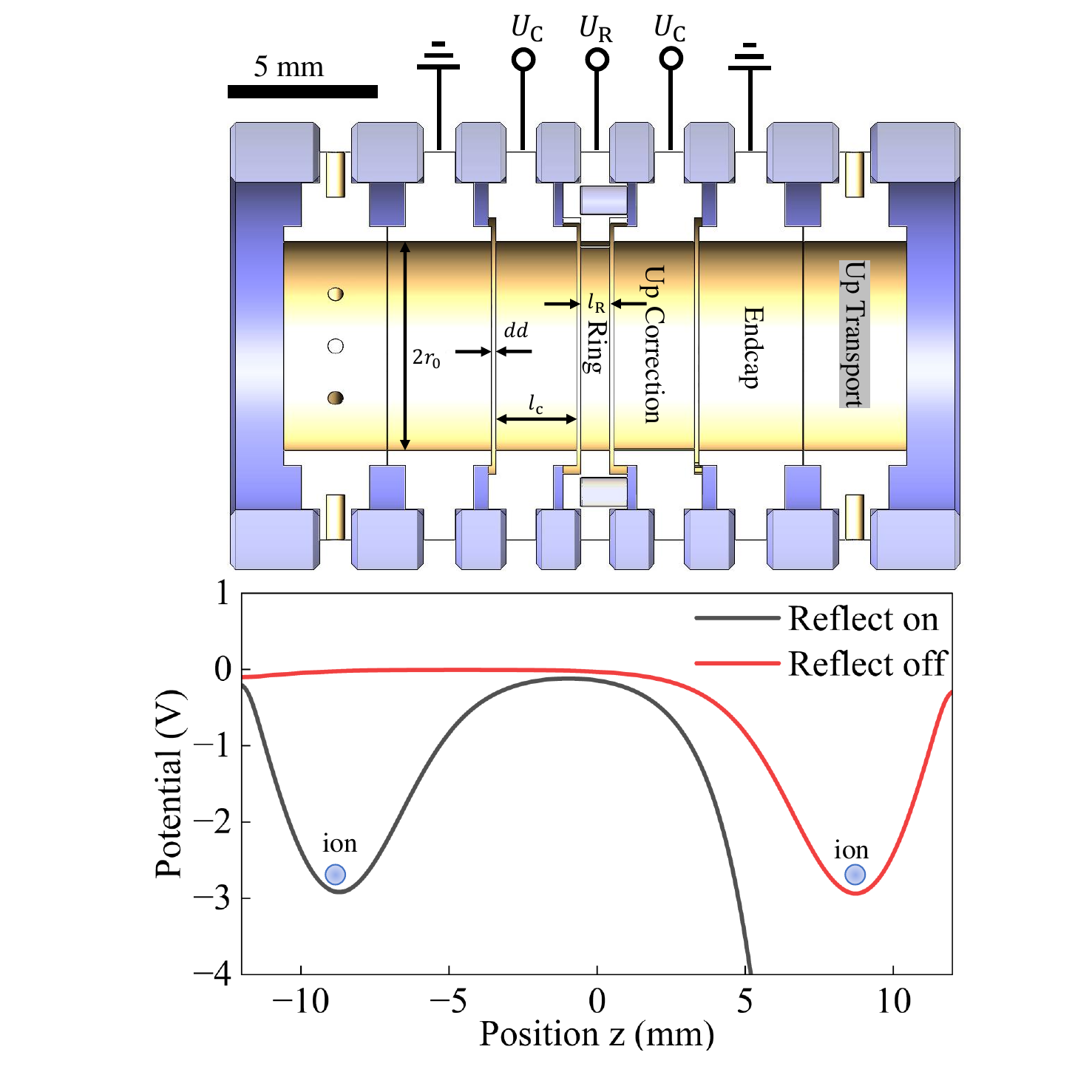}%
\caption{\label{fig_ionG} Vertical cross-section of the five-electrode trap structure with two transport electrodes. The gold-colored components represent gold-plated copper electrodes, which are separated by sapphire insulating rings (blue). The relevant electrode dimensions from Table~\ref{table_parameters} are annotated. Below the cross-section, the plot shows the initial electrostatic potential along the trap axis during electron bombardment of the target electrode. The black curve corresponds to the configuration with a negative voltage applied to the reflector electrode and a trapping field established on the left side of the ring electrode. The red curve shows the potential when the reflector is grounded and a trapping field is configured directly on the right side of the ring electrode.}
\centering
\end{figure}

Highly charged ions were produced using an in-trap electron beam ion source (EBIS). As shown in Figure~\ref{fig_trap}, the electron beam originated from a self-made field emission point (FEP). It was accelerated by the voltage difference between the anode and cathode, passed through the trap region, and reached the collector.There are two targets, made of electrically conductive carbon nanotubes, positioned on either side of the trap. Target I was held at ground potential, while the potential of target II could alternatively be set by the voltage applied to the reflector electrode. Without using the reflector, a portion of the electron beam directly strikes target II, causing atoms to be sputtered from its surface. A potential well was set to trap the ions generated by ionization of electron collisions as shown in Figure~\ref{fig_ionG}. Similarly, when a voltage was applied to the reflector to block the beam, the electrons were reflected back and forth before finally striking Target I, as employed in Ref.~\onlinecite{Sturm2014}, leaving the opposite side available for other applications. In this work, we tested both approaches to produce HCIs. It should be noted that, due to space limitations, the voltage applied to the reflector can penetrate the trap region; therefore, the potential well must be carefully configured to ensure successful ion creation. In our experimental configuration, the cathode emitted an electron beam of 62.8~nA. On the target I and target II, the beam current were measured to be 45.3~nA and 10~nA, respectively, with a typical bias of –800~V on the cathode and –200~V on the extractor. When tuning the reflector voltage to -300~V, the current on target I changed to 52~nA and that of target II reduced to 1.5~nA after a few seconds of beam reflection.  It is worth noting that the majority of electrons were reflected even when the reflection voltage was lower than the extraction potential. This behavior can be attributed to the weaker magnetic field at the FEP compared to that at the reflector, which reduces the axial kinetic energy of electrons due to the magnetic mirror effect. Typically, the electron beam was turned on for more than 30~s, allowing ions to be produced with charge states corresponding to the beam energy.

After creating HCIs in the orignal potential well, they were transported to the ring electrode of the main trap, as shown in Figure~\ref{fig_trap}. The voltages of the trap electrode assembly were optimized as follows: the ring electrode $U_{\mathrm{R}}$ was set to a voltage corresponding to the ions’ charge-to-mass ratio; the two correction electrodes were set to $U_{\mathrm{c}} = T_{\mathrm{r}} \cdot U_{\mathrm{R}}$, where $T_{\mathrm{r}} = 0.8827$ was obtained from a finite element electromagnetic simulation to minimize the higher order harmonic coefficient $C_4$; both endcap electrodes were grounded.

\begin{figure}[!htbp]
\includegraphics[width=0.5\textwidth]{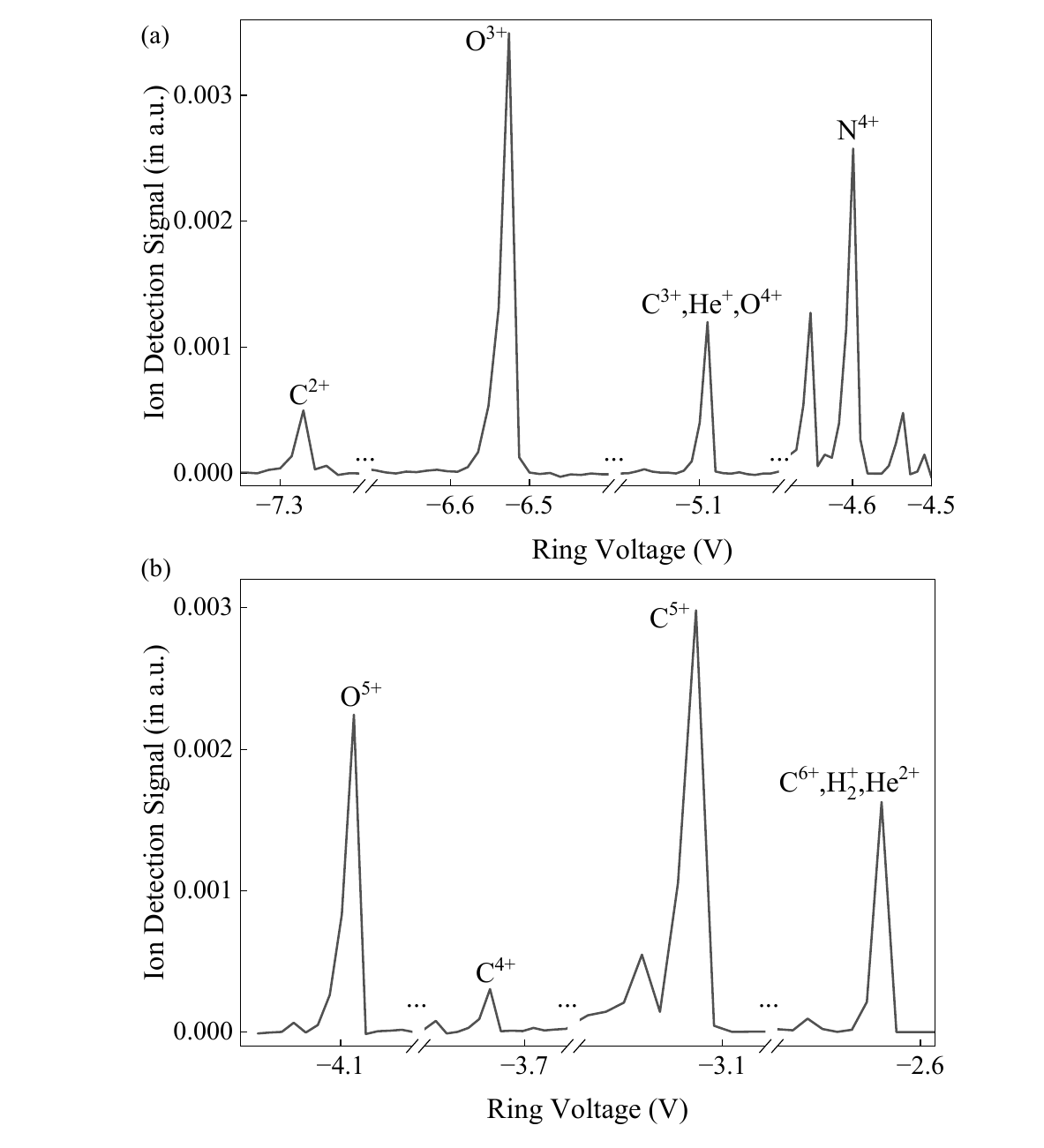}%
\caption{\label{fig_scan} Ring voltage scan spectrum. Panels (a) and (b) together display the ring voltage scan results spanning from -7.5~V to -2.6~V. By varying the voltages applied to the ring and correction electrodes, the axial oscillation frequency of the ions is progressively shifted. The sum of the differences between each spectrum (recorded at a specific ring voltage) and the baseline spectrum is calculated. A peak appears in the plot when the condition $\omega_\mathrm{z} = \omega_\mathrm{res}$ is met for a particular ion. The corresponding ring voltage value can then be substituted into Eq.~\ref{eq:UR} to determine the ion species.}
\centering
\end{figure}

At the beginning of the measurement, detecting a clear ions’ image current signal was not simple. This difficulty arose from two factors. First, due to manufacturing and mounting uncertainties, the actual electric potential deviated from the simulation, leading to significant field anharmonicity or misalignment with the magnetic axis. This resulted in a frequency shift and broadening, thereby smearing out the ion peak signal. Second, electronic noise in the environment could bury the very weak ion signal. Therefore, we employed a parametric drive technique, as used in Ref.~\onlinecite{Bushev2008}, to verify the presence of ions. To determine the ion frequencies, we first searched for ion signals at $\frac{q}{m}=1$ and $\frac{1}{2}~\mathrm{e/u}$ via parametric drive. During the frequency sweep of the ring voltage, when the voltage satisfied $\omega_\mathrm{z} = \omega_\mathrm{res}$, i.e.,
\begin{equation}
U_\mathrm{R} = \omega_\mathrm{res}^2 \frac{d_\mathrm{char}^2}{C_2} \frac{m_\mathrm{ion}}{q_\mathrm{ion}} + \Delta U_\mathrm{R},
\label{eq:UR}
\end{equation}
a 0.1 Vpp axial dipole excitation was applied at a frequency $\nu_\mathrm{exc} = 2\nu_{\mathrm{res}}-150~\mathrm{Hz}$. This results in a peak signal on the left side of the resonator, which becomes observable above the resonator noise, as shown in the spectrum of Figure~\ref{fig_trap}. After confirming the presence of the ions, the corresponding voltage for each ion was recalculated using Eq.~\ref{eq:UR}. By scanning the voltage $U_{\mathrm{R}}$, the axial frequency $\nu_{\mathrm{z}}$ of the trapped ions was shifted sequentially to resonance with the resonator frequency, allowing the monitoring of different ion species. As can be seen in Figure~\ref{fig_trap}, when the oscillation frequency of the hot ions approaches the resonator frequency, the entire spectral signal rises. Therefore, we subtracted the baseline (the red line in the spectrum of Figure~\ref{fig_trap}) from the spectrum at each ring voltage and integrated the difference, resulting in the ring voltage scan spectrum shown in Figure~\ref{fig_scan}. The charge-to-mass ratio of the ions can be derived from the value of $U_{\rm{R}}$, enabling identification of the ion species. The ions observed in the experiment include $\rm{C}^{2+}$ to $\rm{C}^{6+}$ produced by bombarding the target electrode, as well as $\rm{O^{3+}}$ to $\rm{O^{5+}}$, $\rm{He^+}$, $\rm{He^{2+}}$, and $\rm{H_2^+}$ resulting from the ionization of the background gas.

In the analysis of the experimental results, a systematic investigation of ion signal broadening and noise-related issues was conducted. Based on the analysis of frequency-shift effects in non-ideal electromagnetic environments, \cite{KETTER20141}the broadening observed in the ion signals from the current setup can be attributed to three interrelated factors: insufficient radial cooling, interactions among a diverse population of ions of multiple species and charge states, and noise-induced heating. During the initial experimental phase, the trap confined a relatively large and heterogeneous ensemble of ions, leading to significant inter-ion interactions. Furthermore, imperfections in the electromagnetic fields originated from machining tolerances of the electrodes (with a maximum deviation of approximately 0.02 mm) and the inherent inhomogeneity of the permanent magnet employed in the assembly.

The predominant external noise source was identified as mechanical vibration from the helium compressor. To evaluate its impact, comparative measurements were performed with the compressor temporarily deactivated. This provided a stable—though limited—data acquisition window of approximately 30 minutes before the superconducting coil quenched. Consequently, the ability to conduct long-duration signal averaging and to finely adjust the trap anharmonicities was constrained under these operating conditions.

\section{\label{sec:conclusion} CONCLUSION AND OUTLOOK}
In summary, this work presents the development and operation of a cryogenic permanent magnet Penning trap. The complete processes of ion generation, transport, confinement, and detection have been demonstrated. This prototype serves two interconnected purposes: first, as a functional and risk-reducing prototype for the construction of the 7~T superconducting-magnet Shanghai Penning Trap, it validates core ion-handling capabilities within an identical electrode structure and under cryogenic conditions; second, as a versatile and efficient standalone test apparatus, it offers a turn-key, helium-free platform for rapid testing of ion sources, beam transport, and storage schemes. Looking ahead, the design is inherently adaptable—it can be reconfigured for spectroscopic applications, and its magnetic field homogeneity can be enhanced through integration with a compact superconducting magnet system. Thus, this work not only provides essential technical validation for the forthcoming Shanghai Penning Trap but also establishes a flexible experimental platform that contributes to the broader development of compact, high-performance Penning traps for precision measurement studies.

\begin{acknowledgments}
The authors extend their sincere gratitude to Professor Sven Sturm from the Max Planck Institute for Nuclear Physics for the invaluable discussions on techniques. The authors also thank Dr. Yi Jiang for her assistance in target preparation. This work was supported by the National Key R\&D Program of China (No. 2023YFA1606501), the National Natural Science Foundation of China (No. 12474251), Max-Planck Partner Group Project, and the Fudan University Yan Liyuan–EnSiKai Foundation (No. JX240003). 
\end{acknowledgments}

\section*{Data availability}
The data that support the findings of this study are available from the corresponding author upon reasonable request.

\nocite{*}
\bibliography{mybib}% Produces the bibliography via BibTeX.

\end{document}